\documentclass[10pt]{article}
\title{Mathematical Software: Past, Present, and 
Future\thanks{Dedicated to John R. Rice on occasion of his 65th 
birthday.  Contribution of the National Institute of Standards 
and Technology, not subject to copyright.  Mention of commercial 
products in this paper does not imply recommendation or 
endorsement by NIST.  Conversely, omission of a product's name 
does not imply unsuitability for use.  Author's address: 
Mathematical and Computational Sciences Division, Information 
Technology Laboratory, NIST, Stop 8910, 100 Bureau Drive, 
Gaithersburg, MD 20899-8910, USA}}
\author{Ronald F. Boisvert \\
        National Institute of Standards and Technology}
\date{}

\begin{document}
\maketitle

\begin{abstract}
This paper provides some reflections on the field of mathematical 
software on the occasion of John Rice's 65th birthday.  I 
describe some of the common themes of research in this field and 
recall some significant events in its evolution.  Finally, I 
raise a number of issues that are of concern to future developments.
\end{abstract}

\section{Introduction}

\subsection{The study of mathematical software}

The field of mathematical software is concerned with the science 
and engineering of solving mathematical problems with computers.  
The primary focus is the development of general-purpose software 
tools applicable to problems in a variety of disciplines.  There 
are a large number of facets to this work, including the 
following. 

\begin{itemize}

\item the development and analysis of algorithms for standard 
mathematical problems which occur in a wide variety of 
applications

\item the practical implementation of mathematical algorithms on 
computing devices, including study of interactions with 
particular hardware and software systems

\item the environment for the construction of mathematical 
software, such as computer arithmetic systems, languages, and 
related software development tools

\item software design for mathematical computation systems, 
including user interfaces

\item testing and evaluation of mathematical software, including 
methodologies, tools, testbeds, and studies of particular systems

\item issues related to the dissemination and maintenance of 
mathematical software

\end{itemize}

In 1977 John Rice aptly characterized the need for specialized study in 
this area with the following observation \cite{Rice79}.
\begin{quote}\it
Many sophisticated scientists produce naive software just as 
many sophisticated computer programmers produce naive science.
\end{quote}
Tremendous progress has been made in the mathematical software 
field in the past 25 years.  Yet, there continues to be a wide 
range of quality in existing software, in both the research and 
commercial domains.  Good mathematical software results from 
the application of certain principles, methodologies, and 
practices derived from both applied mathematics and computer 
science.  The 
study of these principles and practices is central to the field 
of mathematical software.  To this end, typical software 
engineering practices, while beneficial to the production of 
mathematical software systems, are not sufficient.  
Mathematical software operates in the milieu of scientific 
computing, which has a number of characteristics that distinguish 
it from other areas.  Among these are the following.

\begin{itemize}
\item {\bf Floating-point arithmetic.} Most scientific 
computations are performed with floating-point arithmetic.  
Consequently, rounding errors occur in most arithmetic 
operations.  Mathematical algorithms, therefore, must not only be 
correct in a strict mathematical sense, but they must control the 
accumulation of round-off, avoid catastrophic loss of 
significance from the subtraction of like quantities, and avoid 
unnecessary overflows and underflows.  Such problems are sometimes
unavoidable; software systems must be 
designed so that they do not fail when these anomalies do occur.

\item {\bf Approximations.}  Floating-point arithmetic certainly 
implies approximation at a very fine level.  However, more 
substantial approximations occur in mathematical computation.  
Infinite series are truncated, difficult-to-compute functions are 
approximated by polynomials, derivatives are approximated by 
differences, integrals are approximated by finite sums, curved domains 
are approximated by polygonal ones.  The combined effect of such 
approximations on the final result can be quite difficult to 
assess.  Analysis must be used to show that the correct solution 
is obtained as the approximations are made more precise (i.e., 
that the algorithm is convergent).  However, good software must 
do more.  It must provide mechanisms for a user to assess the 
quality of the result, and to alert the user when the result is 
suspect.  Well engineered software can use such metrics to 
automatically control the level of approximations, optimally 
adapting the algorithm to the situation at hand.

\item {\bf Infinite processes.}  Many mathematical computations 
consist of applying some infinite process that obtains the 
desired result only in the limit.  Such processes must be 
truncated for practical use.  Considerable research efforts have 
been involved in finding iterations that converge quickly.  
Deciding when to stop is always a difficult problem of practical 
concern to software developers.  Good software must employ 
techniques that detect divergence or too slow convergence and 
take appropriate action.
\end{itemize}

Coupled with these fundamental mathematical challenges are 
practical concerns about portability.  How can developers produce 
software with reliable, reproducible behavior when it must run in 
very different environments, with different types of processor 
architectures, arithmetic systems, memory hierarchies, operating 
systems, and language processors?  Such questions are critical in 
the study of mathematical software.

\subsection{The contributions of John R. Rice}
 
At this conference, we are celebrating John Rice's long and 
influential research career.  John has made fundamental 
contributions to the areas of approximation theory, numerical 
analysis, mathematical software, and computer science.  In the 
area of mathematical software, his technical contributions have 
had three overriding themes.

\begin{enumerate}

\item {\bf Architecture of scientific software systems.}  John 
has participated in the design and development of a variety of 
widely distributed mathematical software systems 
\cite{HoMR85, HoMR85a, RiRW84, Rice78, Rice84, RiBo85}.
In the course of this 
work he pioneered a number of design concepts which have 
influenced many systems. Among these are {\em 
polyalgorithms} \cite{Rice68}, {\em meta-algorithms} 
\cite{Rice75}, and {\em software parts} \cite{RiSw87}.

\item {\bf Raising the level of abstraction.}  Software users are 
more efficient when they can express their computational needs in 
the language of their technical field.  For applied mathematics, 
abstractions are based upon concepts of the calculus, not simple 
arithmetic operations encapsulated in programming languages like 
Fortran.  John Rice's work in high-level components and languages 
\cite{RiBo85,RiSw87}, intelligent interfaces 
\cite{HoRV90,HoRV92a,HoRV92}, and problem-solving environments 
\cite{GaHR94,RiBo96} have served to push abstractions to higher 
and higher levels.

\item {\bf Understanding software via experimentation.}  
Understanding the behavior of software is necessary in order to 
make practical decisions regarding algorithm selection
\cite{Rice76}. John has often stressed the importance of the use 
of experimentation in such evaluations.  The many small 
engineering decisions made in the course of translating an 
abstract algorithm into a working computer program can have an 
enormous impact on its performance characteristics.  John has 
devoted much time to developing testing and evaluation 
methodology \cite{BoHR79,HoRi80,RiHD81,Rice79a,WHRJ96}, and 
applying it to particular situations \cite{DyRR88, DHLR84, 
HoLR78, Rice83}.  Indeed, one of the principle applications of 
the ELLPACK system \cite{RiBo85} and its successors has been to 
the performance evaluation of software for partial differential 
equations.

\end{enumerate}

John's contributions to the field of mathematical software have 
been voluminous and far-reaching.  (In this paper I have only 
cited a few examples of his many writings on this subject.) 
In the remainder of this paper I will enumerate some of the major 
events in math software, pointing out some of John's key 
contributions along the way.  I will then describe several 
current issues facing the field and make several hazy predictions 
of the future.

\section{Mathematical Software Past}

\subsection{Beginnings}

The earliest applications of electronic computers were in science 
and engineering, for which mathematical computation played a 
central role.  Programming was a very difficult chore, done 
without modern aids like high-level languages, compilers and 
debuggers.  The first publication of a piece of mathematical 
software in a research journal probably occurred in 1949, when {\it Mathematical Tables 
and Other Aids to Computation} printed a UNIVAC code for the 
solution of Laplace's equation written in machine language \cite{SnLi49}.  
Such codes were very difficult to produce, and the need for reuse of 
software was recognized very early on.  In 1951, Wilkes, Wheeler 
and Gill presented one of the earliest program libraries, which 
was developed for the EDSAC\footnote{The EDSAC (Electronic Delay 
Storage Automatic Computer) was built in the late 1940s at the
Mathematical Laboratory of the University of Cambridge. It was
operational from 1949 until 1958.} \cite{WiWG51}.  

By the 1960s, the introduction of high-level 
programming languages, e.g. Algol and Fortran, had greatly eased 
the task of producing reusable mathematical software. The use of 
such languages was not without controversy, of course. Compiled 
code was not quite as efficient as hand-tuned assembly code, but 
most people were willing to accept this in light of the great 
savings in programmer time.  Also, the subprogram structure 
provided by these languages provided a simple framework for the 
construction and maintenance of libraries of utilities. 

In 1960, the Association for Computing Machinery (ACM) began a new
editorial department in the {\it Communications of the ACM} (CACM)
devoted to the publishing of algorithms.
Edited by J.~H.~Wegstein of the National Bureau of Standards (NBS),
this section printed the code of contributed Algol procedures
(most such codes were quite short).  Also, remarks on and
certifications of previously published codes were solicited. 
The first such contribution was a code for numerical quadrature 
submitted by R.~J.~Herbold of NBS \cite{Herb60}.  Each algorithm 
was given a number, and the set of algorithms later became known 
as the {\em Collected Algorithms of the ACM} (CALGO). 

Computer manufacturers also began to develop libraries for their 
users.  The most prominent of these was probably the IBM 
Scientific Software Package (SSP).  A number of laboratories, 
such as Bell Labs, Boeing, Harwell, and Monsanto, began the 
development of math software libraries for their internal use.
Several organizations, such as SHARE, the IBM user's group, began 
to collect such utilities for redistribution.  

Of course, subroutine libraries were not the only focus of 
researchers in this new field.  Some were imagining ways in which
these new powerful computers could be used to transform the way
in which applied mathematics was practiced.
Many of these ideas were discussed at the Symposium on Interactive Systems 
for Experimental Applied Mathematics held in in Washington, D.C. in August 1967
\cite{KlRe68}.  The vision there is remarkably clear; 
many of the participants reported on developments in technologies which would 
only finally begin to be realized in the 1980s and 1990s.  
At Purdue, for example, John Rice, Saul Rosen and colleagues
designed NAPSS (Numerical Analysis Problem-Solving System), an interactive 
mathematical problem-solving system which would accept input akin 
to normal mathematical notation \cite{RiRo66}, and would employ a variety
of heuristics to automate numerical analysis.  Unfortunately, the resources 
necessary for such an ambitious system exceeded even the 
supercomputers of the day (like the CDC 6400), and a fully 
functional system was never realized.

\subsection{A community emerges}

Perhaps the first event that provided a real sense of community for 
researchers interested in the production and dissemination 
of reusable mathematical software 
was the {\em Mathematical Software Symposium} held at Purdue 
University in April 1970. John Rice organized the 
symposium\footnote{The organizing committee included Robert 
Ashenhurst, Charles Lawson, M. Stuart Lynn, and Joseph Traub}, which 
was sponsored by ACM and the Office of Naval Research, 
and the proceedings were published 
as a book by Academic Press \cite{Rice71}. Included in the 
proceedings were 23 papers, four descriptions of selected 
mathematical software, and more than 40 pages of introductory 
material prepared by Rice.

One of the recommendations from the Symposium was for the 
establishment of a journal that would publish papers related to 
mathematical software.  John Rice vigorously pursued this 
possibility\footnote{A committee that included Wayne Cowell, 
Lloyd Fosdick, Tom Hull, M. Stuart Lynn, and Joseph Traub worked 
with him.}.  After considerable negotiations with ACM and the
Society for Industrial and Applied Mathematics (SIAM), 
ACM agreed to publish the new journal.  Papers from an NSF-sponsored 
conference were used to provide articles to 
seed the journal.  {\em Mathematical Software II} was held at 
Purdue in May 1974.  There were 225 attendees, with 82 papers 
presented.  The best of those papers make up the majority of the
first volume of the {\em ACM Transactions on Mathematical Software} 
(TOMS) which published its first issue in 1975 with John Rice as 
Editor-in-Chief.  John continued in that position until 1993.

TOMS was chartered not only to publish traditional research 
papers, but also algorithms (with included code which would be 
refereed), certifications, translations, and remarks on 
previously published algorithms.  The Algorithms section of CACM 
was moved to TOMS, and hence TOMS 
algorithms were numbered beginning at 493.  One of the important 
features of the new journal was the establishment of a reliable 
Algorithms Distribution Service for CALGO.  The distribution, on 
magnetic tape, was performed on a cost-recovery basis by IMSL, 
Inc. Obtaining software in machine-readable form was much more 
useful than reading code on paper. This also allowed TOMS to 
adopt the policy of not printing the code of algorithms in the 
pages of its journal, thus saving much in production costs.

A third conference organized by John Rice, {\em Mathematical 
Software III}, was held at University of Wisconsin in 1977 
\cite{Rice77}.  By the end of the decade, mathematical software 
had emerged as a viable research area with an active community 
to support it.  After publishing 25 volumes, TOMS remains a 
vibrant outlet for the work of this community \cite{TOMS00}.

\subsection{Software emerges}

Another important activity in the 1970s were the numerous efforts 
to develop carefully constructed, systematized collections of 
mathematical software. One of the first of these was the NATS 
project, the National Activity to Test Software, which was 
conceived in 1970.  A joint venture of Argonne National 
Laboratory, Stanford University, and the University of Texas at 
Austin, NATS was designed to study problems in producing, 
certifying, distributing, and maintaining quality numerical 
software.  A key part of this effort was the production of two 
Fortran software packages, EISPACK \cite{GBDM77,SBDG76} for 
eigenvalue problems, and FUNPACK \cite{Cody75} for special functions. 

EISPACK, which first appeared in 1972, was based upon algorithms 
published in the 1960s in {\em Numerische Mathematik} and later 
collected by Wilkinson and Reinsch in the {\em Handbook for 
Automatic Computation} \cite{WiRe71}.  Although the core of 
EISPACK was largely a Fortran translation of these existing Algol 
codes, the project was enormously influential.  It set a new 
standard for quality transportable mathematical software, 
rigorously tested in a wide variety of computing environments. 
Its success inspired the development of many systematized 
collections, or ``PACKs'', in other areas: 
LINPACK for linear systems \cite{DBMS79}, 
FISHPAK for separable elliptic problems \cite{SwSw79}, 
DeBoor's B-spline interpolation package \cite{deBo79}, 
MINPACK for nonlinear systems \cite{MoGH80},
DEPAC for ordinary differential equations \cite{ShWa80},
Fullerton's function library FNLIB \cite{Full77}, 
Swarztrauber's FFTPACK for fast Fourier transforms \cite{Swar82},
and QUADPACK for numerical quadrature \cite{PiDK83}.

Of all the early ``PACKs'', LINPACK undoubtably saw the most 
widespread use.  One of the keys to LINPACK's success was the 
decision to base its coding on the newly proposed Basic Linear 
Algebra Subprograms (BLAS) \cite{LHKK79}. The BLAS performed 
elementary vector operations, such as norms, dot products, 
scaling, and vector sums.  The innermost loops in LINPACK's 
column-oriented algorithms occurred inside the BLAS.  This 
allowed optimization of the whole package by simply optimizing 
the BLAS. This approach proved quite successful, and many 
machine-specific versions of the BLAS were developed and 
supported by computer manufacturers.

The 1970s also saw great advances in software for ordinary 
differential equations (ODEs). Gear's code DIFSUB \cite{Gear71} 
provided a well designed framework for automatic integration of both 
stiff and non-stiff problems using linear multistep methods. 
Shampine and Gordon's 
ODE \cite{ShGo75} did the same for Runge-Kutta methods. Many 
subsequent packages were built using these basic designs.  Other 
influential packages included COLSYS for two-point boundary-value 
problems \cite{AsCR79} and DASSL\footnote{DASSL won the 1991 
Wilkinson Prize for Numerical Software} for differential 
algebraic systems \cite{Petz82}.

A number of high quality multi-purpose libraries also had their 
start in the 1970s.  In 1970, six British computing centers began 
an effort to develop a library for their ICL 1906A/S computers.  
The next year Mark 1 of the Nottingham Algorithm's 
Group (NAG) library was released.  Implementations for other 
systems followed, and in 1976 a not-for-profit company, Numerical 
Algorithms Group, Ltd., was formed to continue development and 
distribution.  The NAG effort continues today \cite{NAG00}.  The 
first commercial math library effort was also begun in 1970 with 
the incorporation in Houston, Texas of IMSL, Inc. by Charles W. 
Johnson and Edward Battiste. By the time the company showed its 
first profit in 1976, there were 430 library subscribers; the IMSL 
library remains a viable commercial product \cite{IMSL00}.
Bell Laboratories also developed a library, PORT, whose single-source 
approach to portability influenced many subsequent efforts
\cite{FoHS78}.

The development of ELLPACK, a system for elliptic boundary-value 
problems, also began in the mid 1970s.  This effort, which was 
led by John Rice, was a cooperative project of Purdue University, 
the University of Texas at Austin, Yale University, and others.  
In ELLPACK, the solution process was partitioned into a number of 
distinct phases (domain processing, discretization, indexing, 
linear system solution, and output), and the interfaces between 
these phases were carefully defined. This design allowed the 
development of a large library of components which could be 
easily composed to build algorithms for solving particular 
problems. ELLPACK also proved to be an excellent testbed for the 
evaluation of software for elliptic problems. To ease the use of 
the system, John Rice designed a very-high-level language to describe 
the problem to be solved, and to select the components to be used 
to solve it.  The system first became fully operational in 1978 
\cite{RiBo85}.  Many of the basic concepts in ELLPACK's design, 
such as high-level user interfaces and plug-and-play software 
parts technology, are in common use in modern problem-solving 
environments.

The development of mathematical software in the 1970s and early 1980s
is described in detail in the book {\em Sources and Development of
Mathematical Software} edited by Wayne Cowell \cite{Cowe84}.

\subsection{Increased access}

By the beginning of the 1980s a substantial collection of 
mathematical software, mostly in the form of Fortran subprograms, 
was available for use.  The user base for this software had grown 
substantially, and with it came a new problem: how to locate that 
needed software component.  The National Bureau of Standards (now 
NIST) developed an extensive catalog of such software. Their 
Guide to Available Mathematical Software (GAMS), based upon a 
detailed tree-structured problem-oriented classification system \cite{BoHK85}, 
allowed readers to see which components of which libraries and 
packages, both public domain and commercial, were available to 
solve each problem.  The catalog remains available today as an 
online resource \cite{NIST00}.

Another barrier to the widespread use of software developed by 
the research community was simply the process of obtaining the 
code.  One had to locate the author, request a magnetic tape, 
and attempt to decipher its format.  Useful software was often 
lost to the community when an author changed institutions and 
there was no longer support for distributing it.  In 1985, Jack 
Dongarra, then at Argonne National Labs, and Eric Grosse at Bell 
Labs, started a software repository they called {\em netlib} 
\cite{DoGr87}, which pioneered the use of computer networks in 
software distribution.  Software could be obtained automatically
by return email after 
sending requests to an address whose email was processed by a Unix 
daemon.  The ready availability of such software changed the way 
in which many researchers worked.  Many more made routine use of 
high quality software, and many others were freed of the 
necessity of maintaining their own private repositories.  Now 
Web-accessible and supported by the University of Tennessee at 
Knoxville and Bell Labs, with mirrors worldwide, {\em netlib} 
remains the premier repository of software developed by the 
mathematical software community \cite{BDGR95}.

The 1980s also saw the first commercial success for general-purpose 
interactive systems for mathematics.  A system for matrix 
computations developed as a teaching aid during the period 1977-84
by Cleve Moler at the University of New Mexico, was 
commercialized as a tool for control system engineers.  Today MATLAB is 
a very popular system for scientific computing \cite{Matl00}.  
The overall structure of modern 
interactive mathematics systems was greatly influenced by the 
system Mathematica developed by Stephen Wolfram in 1988 
\cite{Wolf99}.  Mathematica was the first commercial system to 
integrate symbolic, numerical, and graphical capabilities into a 
single package.  The growing availability of personal computers and 
workstations was an important factor in the success of these 
systems.  With these tools, the use of mathematical software was 
beginning to expand to those with little experience in numerical 
methods or even programming.

\subsection{New architectures}

The 1980s also brought vector and parallel computers into widespread 
use, and with them additional challenges to the design of 
mathematical software.  Vector processor vendors developed 
specialized math libraries tuned for their systems, mainly 
containing software for linear systems and FFTs.  These solutions 
emerged because the performance of linear algebra software such 
as LINPACK was disappointing on vector register architectures 
like the Cray and Convex.  The main reason for this was the fact 
that moving data from memory to vector registers was very costly, 
and that LINPACK's column-oriented algorithms, based on the BLAS, 
necessitated more data movement than was really necessary. 

In 1984 John Rice hosted a workshop at Purdue (``ParVec 
Workshop Number 4'') in which a variety of schemes for developing portable 
high-performance software for vector parallel systems were 
proposed \cite{Rice84a}.  
Jack Dongarra and Sven Hammarling proposed the development of 
new classes of BLAS:  
Level 2 BLAS for matrix-vector operations, and Level 3 BLAS for 
matrix-matrix operations.  Encapsulating $O(n^2)$ and $O(n^3)$ 
operations, respectively, as fundamental operations would provide 
much more opportunity to optimize core operations on different 
processors.  These new BLAS \cite{DDHD90,DDHH88} would provide 
the basis for a major new linear algebra package released in 
1992. LAPACK \cite{ABBB99}, which included the functionality of 
both EISPACK and LINPACK, used block-oriented algorithms 
in which the fundamental operations were now matrix operations 
encapsulated in the Level 3 BLAS.  These have proven to be highly 
efficient on modern vector processors and symmetric multiprocessors.  
Every major computer manufacturer now supports tuned BLAS for 
their systems and incorporates LAPACK in their math library.  
Community efforts are currently underway to extend the BLAS in 
new directions, such as sparse matrix operations \cite{BLAST00}

In the late 1980s multiprocessor systems of widely differing design 
were becoming routinely available, and with them a host of new 
programming models, supported by specialized message-passing 
primitives.  Developing portable software for the class of
distributed memory (multiple instruction multiple data, or MIMD) 
systems became a new challenge.  The PVM system developed  
in 1991 provided a useful abstraction 
for parallel programming and was very widely adopted 
\cite{GBDJ94}.  Its implementation on many parallel machines 
demonstrated the usefulness and feasibility of a common 
message-passing infrastructure.  This led to a grass root message-passing 
standardization effort.  The resulting Message Passing Interface 
(MPI) transformed the landscape for distributed parallel 
computing \cite{GSNL98}.

One of the first portable math software libraries for distributed 
architectures was ScaLAPACK, a distributed memory counterpart of 
LAPACK linear system solvers \cite{BCCD97}.  This package became 
the core of several multi-purpose distributed memory math 
software libraries which first appeared in the 1990s.  Among 
these are the NAG Parallel Library \cite{NAG00}, IBM's PESSL 
\cite{IBM97}, and the European PINEAPL effort \cite{PINE00}.  

The increasing complexity of scientific software systems being 
developed in the 1990s led to an interest in new software 
architectures.  Object-oriented approaches to the development of 
mathematical software began to be seriously considered.  The 
notorious inefficiencies of pure object-oriented design, and the 
lack of language standardization made such pursuits difficult.  
Nevertheless, approaches that allowed many of the advantages of 
object-oriented design without sacrificing efficiency were 
developed.  LAPACK++, a subset of the linear systems solvers in 
LAPACK written in C++, was one of the first such successful 
packages \cite{DoPW93}.  Today object-oriented approaches are 
routinely used in scientific computing.

\subsection{Expanding vision}

By the 1990s, rapidly increasing computer power was leading to new visions for 
the future of mathematical software systems.  During that period, 
for example, John Rice and colleagues led in the establishment of 
a new community of researchers interested in exploiting the 
promise of expert systems for numerical computing.  In a series 
of conferences held at Purdue \cite{HoRV90,HoRV92a,HoRV92}, the 
use of AI approaches for such tasks as algorithm selection, 
automatic programming, and process management were explored.

By this time, computation had become an essential 
ingredient in the practice of science and engineering.  Interest 
in computational science as a new field of study was beginning, 
and interdisciplinary programs for training its practitioners 
were being established in many universities.  John Rice and 
others began to develop a new vision for mathematical software 
systems to support computational science research.  These 
systems, called {\em problem-solving environments} \cite{GaHR94, 
RiBo96}, would provide natural graphical user interfaces in which 
scientists describe their problems using the vocabulary of 
their native discipline.  They would provide access to rich 
libraries of problem-solving components enabling Web-based 
parallel and distributed computation.  Users would be able to 
interact with ongoing computations, to easily visualize results, 
to manage a large database of experimental results, and to ask 
advice of expert advisory systems.  Many small-scale special-purpose 
systems now under development and use can be classified as 
problem-solving environments, and research groups throughout the 
world are working on infrastructure necessary for the routine 
construction and use of such systems.  Work at Purdue on parallel 
ELLPACK \cite{HoRP89}, Web\-ELLPACK \cite{MWHR97}, and PYTHIA 
\cite{WHRJ96} are serving to address issues in PSE design.  
Examples of current work in network-based scientific computing
are Netsolve \cite{CaDo98}, the NEOS optimization server
\cite{CzMM98}, and the computational grid \cite{FoKe97}.

The vision of scientific computation in heterogeneous distributed 
environments places stringent requirements on the portability and 
interoperation of scientific software that are extremely 
difficult to achieve \cite{BCPW97}.  Such needs have sparked interest 
in the use of common virtual environments such as Java\footnote{Java is a 
trademark of Sun Microsystems.} for 
computational science and engineering.  The Java language and its 
environment (the Java Virtual Machine), which has become 
available on nearly every computing platform, provides a fixed 
floating-point model, threads, remote execution, standard GUIs, 
and other facilities within a simple object-oriented programming 
language.  While these are the main facilities necessary for the 
construction of problem-solving environments, there has been some 
reluctance to adopt Java within the scientific community due to 
concerns about efficiency and the lack of several programming 
conveniences important to scientists and engineers \cite{BDPR98}.  
Community efforts such as the Java Grande Forum are seeking to
improve this situation \cite{JGFN00}.

Virtual environments do not necessarily solve the problem of 
performance portability, since virtual machine instructions must 
be mapped on to local computer hardware for execution.  Modern 
computing hardware is extremely complex, characterized by 
multiple processing units, vector pipes, register farms, several 
levels of cache (with increasing access times), local memory, 
remote memory, and disk storage.  Getting the highest performance 
possible requires that the programmer take into account all the 
special properties of the system in use.  This leads to extremely 
complex software even for the simplest of tasks.  Matrix 
multiplication can turn into a 10,000-line polyalgorithm.  Recent 
approaches have provided new hope for overcoming this 
software development nightmare.  Clint Whaley and Jack Dongarra 
have recently developed a system, called ATLAS, for Automatically 
Tuned Linear Algebra Subprograms \cite{WaDo98}.  ATLAS generates highly 
efficient BLAS for a given architecture using an experimental 
approach.  By running many hundreds of tests, ATLAS determines 
the most efficient way to implement a given operation.  The 
result is consistently on par with, and often exceeding, code 
which takes expert programmers weeks to develop.  
Matteo Frigo and Stephen Johnson have taken 
a similar approach in the computation of fast Fourier transform 
\cite{FrJo98}.  For FFTs, hardware also interacts with the prime 
factorization of the sequence length $n$ to add further 
complication.  FFTW, the Fastest Fourier Transform in the West, 
uses heuristics and experimentation to develop a just-in-time 
strategy for fast computation for a given $n$ on a given 
processor.\footnote{FFTW won the 1999 Wilkinson Prize for 
numerical software.}

\section{ Mathematical Software Present}

In this section I point out a variety of meta-issues that face 
mathematical software researchers today.

\subsection{Mass-market software}

Until recently mathematical software was produced mostly by experts in 
numerical analysis as a byproduct of their research in algorithms.
Users of this software also were fairly sophisticated, with some 
experience in numerical
algorithm development themselves.  They had an appreciation of the
limitations of numerical algorithms, and the necessity of careful
verification of results, even when using software developed by experts.

Today's community of mathematical software developers and users is much
larger, and much more diverse.  The great demand for mathematical
computations has made mass-marketed mathematical software
profitable.  Commercially supported
mathematical and statistical software is now widely available, with high-level
interfaces that allow use by non-programmers.  Such users often do not have
the background necessary to recognize the difference between a difficult
problem and a routine one.  The mathematical landscape is still littered 
with pitfalls, and these users may be too trusting of the results produced 
by the scientific software systems that they use.  Programmers who add
mathematical and statistical capabilities to commercial software systems 
are no longer experts in numerical analysis.  They may be content to code
up a formula from a book without giving thought to its numerical properties.
The problem may be even more severe in systems that are not overtly
mathematical in nature.  Mathematical computations are increasingly being
done in embedded devices, coded by programmers whose mathematical
sophistication may be suspect.

As a result, in spite of tremendous progress in numerical methods and 
software, many users of modern mathematical software are at risk.  There is 
now a desperate need for numerical analysts to develop and apply
methodologies for the validation of mathematical and statistical software.
Techniques, tools, reference data, and reference software are needed to 
support critical evaluations of mathematical software by developers and
users \cite{Bois97}.  
Unfortunately, there is little interest and support within the 
research community for such activities.  

Those software developers who seek advice regarding numerical software
production are likely to look to popular sources like Numerical Recipes
\cite{PTVF93}.  Books like this provide a reasonably good introduction to
numerical methods, and the programs they include provide good examples of
the basic techniques.  Programs like these are often incorporated
wholesale into applications, in spite of the fact that they are typically
less efficient, robust, and reliable than state-of-the-art mathematical
software.  The mathematical software community needs more popularizers
who can bring the message of good numerical software design to those in
other fields.

\subsection{Tower of Babel}

For many years there was one language for scientific computing: Fortran.
This greatly simplified the development and reuse of mathematical software
components.  Today we are faced with a plethora of programming languages
in use for scientific computing.  Though officially obsolete, Fortran 77 is 
still the language of
many.  Good compilers are now available for Fortran 90, and many have  
been extended to support Fortran 95, the current Fortran
standard, although their adoption by programmers has been slow in coming.  The C language
has proven much more popular, for which excellent compilers are now
available.  Most GAMS users who cannot find the software they seek are 
looking for C procedures.  C++, the object-oriented extension to C, is the
choice for a growing number of new mathematical modeling projects.
Unfortunately, C++ has not, until recently, had an agreed-to standard, and,
as a result, developing portable software has been difficult.  Java, the
popular network-aware object-oriented programming language developed by Sun, 
is being seriously considered by many, although its performance and language
features leave much to be desired.  The fact that Java is now being widely
taught in universities insures its future.  Finally, many software components
are being developed in the very-high-level languages used in specialized 
systems; MATLAB is the primary example.

We are clearly facing a transition in computer languages for science and
engineering computation.  Numerical analysts no longer have much influence
on the choice of language of those doing numerical computing.  
Language choices are more often made based
on other considerations, such as the need for convenient and portable
graphical user interfaces, visualization tools, and other critical
system services.  While such services are largely
unavailable to Fortran programmers wishing to develop portable systems, 
they are conveniently at hand in C,
C++, and Java.  The increased portability afforded by the widespread
availability of Java Virtual Machines on Windows, Unix, and Apple 
platforms, has made Java a very attractive option.  While
mixed-language programming is possible, and does provide the ability 
to reuse legacy Fortran software, this option is not popular among users.
It adds complexity to the software project, while making the code more
difficult to transport.  

Unfortunately, there is almost no support for the migration of the existing base
of Fortran mathematical software components to other languages.
As a result, this well-engineered software is being increasingly bypassed
in favor of inferior home-grown solutions.

\subsection{The risks of self-publishing}

The rise of the Internet has greatly eased the exchange of information
among researchers.  It is simple and convenient for research groups
to develop a Web page to distribute software and documentation to potential
users.  While this has led to increased access to research software, it
places the long-term maintenance of the output of the research community
in jeopardy.  Project Web pages on departmental servers are not permanent
fixtures.  Nevertheless, many researchers are using such facilities in
place of submitting their software to more permanent archives such as
{\em netlib} or the {\em Collected Algorithms of the ACM}.  There
is a danger that much of the currently available expertise embedded in such
software will be lost to future researchers.

\section{Mathematical Software Future}

In this section, I offer a few predictions regarding future
mathematical software research and usage.

\vspace{1em}

\noindent
{\bf Prediction:} Within five years ACM will cease print publication of TOMS.

\vspace{0.75em}

Subscriptions to TOMS have been dropping at the rate of about
5 percent per year for some time.  Other ACM journals, and indeed
most other mathematics and computer science journals, are
experiencing the same phenomenon.  One of the reasons for this is
the proliferation of specialized journals, principally developed
by commercial publishers.  

For some time, ACM has been considering mechanisms for maintaining
their publication program as a viable service to the community.
The solution to this problem is found in the ACM Digital Library
(ACMDL) which premiered in 1998 \cite{ACM00}.  The ACMDL provides
its subscribers with online access to {\em all} ACM journal articles
and conference proceedings published since 1985 at a subscription
fee which is less than the cost of three printed journals.  Currently
this accounts for more than 350,000 pages of text.  Acceptance of the 
ACMDL by members and subscribers has been overwhelming, providing ACM
with the additonal revenue to begin the work of extending the ACMDL 
holdings to include 
all material published by ACM since its inception in 1947.  
At the same time, the success of the ACMDL has contributed to 
a further 25 percent drop in print subscriptions in 1999.  If present
trends continue, printed versions of ACM journals will be no longer
be sustainable in five years time.  Instead, they will be superceded
by their electronic counterparts.

The ACMDL will serve to blur the distinctions between individual ACM journals.
The concept of a journal will be replaced by that of an input stream to the
ACMDL controlled by a certification authority, i.e., a board of editors
supported by volunteer referees.  In such an environment, it will be much
easier (and much less financially risky) for ACM to initiate new refereed
input streams, and to phase out those which have become less productive.
Rather than subscribe to individual journals, ACMDL subscribers will have
access to an individually tailored notification service which will alert
them to the availability of new articles in their areas of interest.

Publications in the ACMDL will not be restricted to articles with a severe
page limit.  Extended appendices will be easily accommodated, as will other
artifacts such as software, audio, video, etc.

\vspace{1em}

\noindent
{\bf Prediction:} Users will no longer install mathematical software on their 
workstations.

\vspace{0.75em}

The need for instantaneous distribution and use of mathematical software 
components in heterogenous network environments will
put new pressures for software portability.  A key element of the solution will
be standardized virtual environments in which software can execute.  Java is
an example of such an environment.  Its widespread availability also provides a 
new model for software distribution.  Complex conglomerations of source code 
will no longer need to be explicitly downloaded and installed on the local 
systems in advance of their use.  Instead pre-compiled bytecodes for the virtual
machine will be able to be downloaded from sites of developers or vendors on
demand.  This also provides a solution to the problem of distributing patches
and updates to software.  Rather than purchasing an entire library, software
users will have the option of subscribing to a service, paying only for the
portions of the library that they actually use.

Another new model for software reuse in a network environment is based on a
remote execution paradigm.  In this case, problem-solving services are made
available to users over the network.  When a problem need be solved, a 
message containing a high-level specification of the problem is sent to
the service provider, who provides both the software and the execution cycles
needed to solve it.  This model is probably more appropriate for access to
large scale systems, like finite-element modeling packages.

\vspace{1em}

\noindent
{\bf Prediction:} The percentage of people directly using math software 
libraries will decrease.

\vspace{0.75em}

The wide availability of problem-solving environments (PSEs) for various domains
will bring computational capabilities to an even wider audience than today.
These users will make use of the services of the PSE, blissfully unaware
of the complex system, involving software libraries, expert systems, and remote 
execution, which are being marshalled on their behalf.

However, if this vision is to be realized, a new class of software designers 
must be trained.  They must be well-versed in numerical analysis, mathematical
algorithms, modern software design, and high-performance computing and
communications.  Additional research in mathematical software must be 
performed to provide new methods for improving the robustness and
adaptibility of mathematical software systems, and to address new problem 
areas.  And finally, new methods
for assessing the correctness and reliability of complex mathematical software
systems must be devised and deployed.  

Mathematical software is still a vital and vibrant research area that will
increase in importance in the coming decades.  We are grateful to John Rice for
his vision and leadership in getting us here.


\end{document}